\documentclass{article}
\usepackage{natbib}
\usepackage[english]{babel}
\usepackage[utf8]{inputenc}
\usepackage{amsmath}
\usepackage{graphicx,hyperref,subfig}
\usepackage{amsmath,amsxtra,amsfonts,amscd,amssymb,bm,amsthm,svg,url}
\usepackage{tikz-cd}
\usepackage{mathrsfs}
\usepackage[colorinlistoftodos]{todonotes}
\usepackage{enumitem}
\usepackage{yfonts}
\usepackage{setspace}
\usepackage{amsfonts,pifont}
\usepackage{calrsfs}
\usepackage[mathscr]{euscript}
\usepackage{geometry}

\usepackage{multirow}
\usepackage[normalem]{ulem}
\useunder{\uline}{\ul}{}

\begin{document}

\title{\textbf{The Knowledge Graph for Macroeconomic Analysis with Alternative Big Data}}

\author{Yucheng Yang\hspace{6mm}  Yue Pang\hspace{6mm}  Guanhua Huang \hspace{6mm}  Weinan E\thanks{Yucheng Yang and Weinan E: Princeton University. Yue Pang: Peking University. Guanhua Huang: University of Science and Technology China. We thank Feng Lu, Chris Sims, Yi Zhang, Lei Zou and audience in the BIBDR Economics and Big Data Workshop for helpful comments, and thank Hanrong Liu, Tao Wen and Lu Yang for research assistance. All errors are our own. Correspondence should be addressed to Yucheng Yang at \url{yuchengy@princeton.edu} and Weinan E at \url{weinan@princeton.edu}.}}
\date{\vspace{3mm}October 2020}
\maketitle
\vspace{5mm}
\doublespacing
\noindent

\begin{abstract}
The current knowledge system of macroeconomics is built on interactions among a small number of variables, since traditional macroeconomic models can mostly handle a handful of inputs. Recent work using big data suggests that a much larger number of variables are active in driving the dynamics of the aggregate economy.  In this paper, we introduce a knowledge graph (KG) that consists of not only linkages between traditional economic variables but also new alternative big data variables. We extract these new variables and the linkages by applying advanced natural language processing (NLP) tools on the massive textual data of academic literature and research reports.  As one example of the potential applications, we use it as the prior knowledge to select variables for economic forecasting models in macroeconomics. Compared to statistical variable selection methods, KG-based methods achieve significantly higher forecasting accuracy, especially for long run forecasts.  
\end{abstract}
\vspace{10mm}
\vspace{10mm}

\newpage
\section{Introduction}
 Traditional macroeconomic models, whether statistical models like VAR \citep{sims1972money} or structural models like DSGE \citep*{christiano2005nominal,smets2007shocks}, can only handle a handful of variables. 
At the same time, the whole knowledge system of macroeconomics is built on our understanding of the interactions among these small number of variables. With the rise of big
 data and machine learning, we now have the opportunity to develop more sophisticated models with a much larger number of variables \citep{mccracken2016fred, coulombe2019machine}.
  In order to do this effectively, a new knowledge system is needed to describe both the statistical and structural relationships of the
  traditional as well as the many new  economic variables. \\

\begin{figure}[h!]
\begin{center}
  \subfloat[Knowledge Graph used by Google Search Engine]{\includegraphics[width=0.45\textwidth]{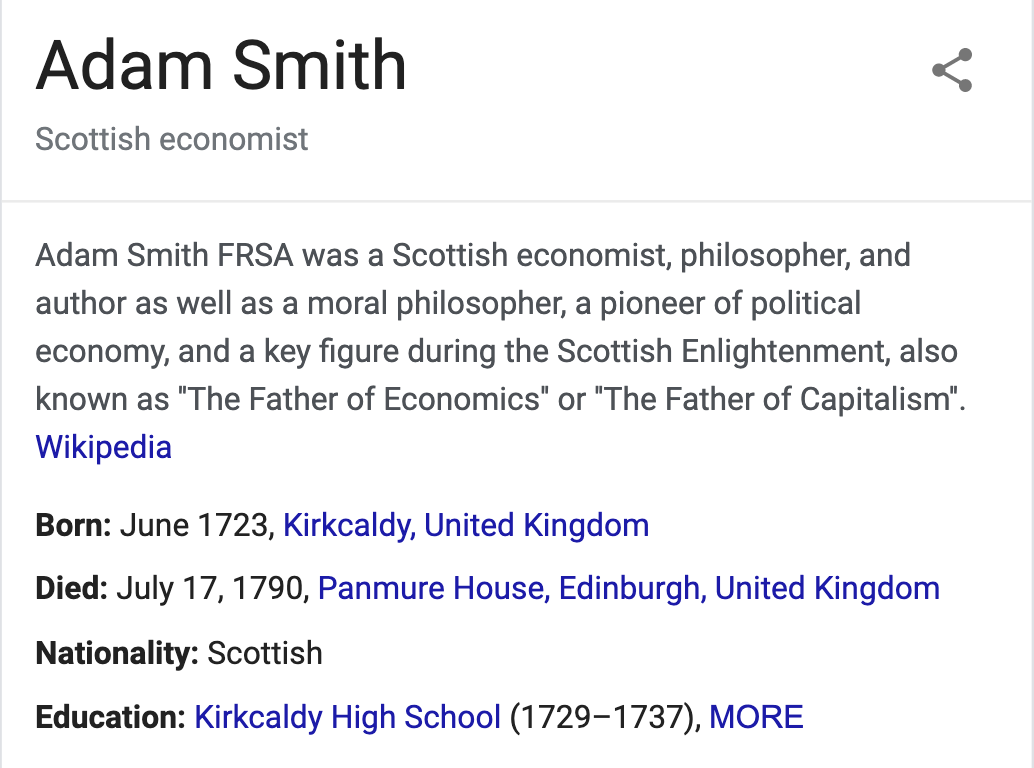}}
  \hfill
  \subfloat[Knowledge Graph of Economic Variables]{\includegraphics[width=0.54\textwidth]{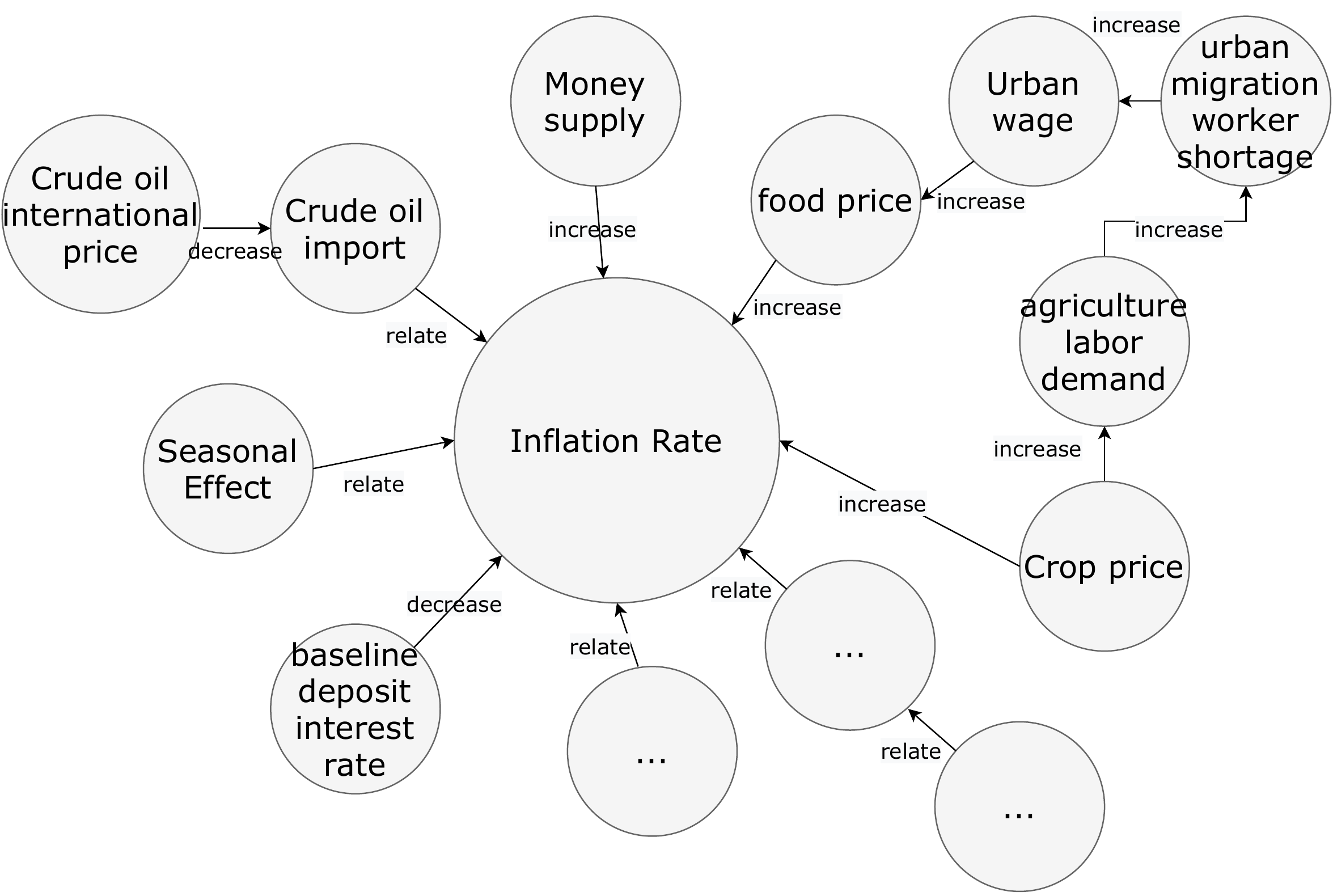}}
  \caption{Examples of Knowledge Graphs}
    \label{fig:kg_ex}
  \end{center}
\end{figure}
In this paper, we discuss how to build such a knowledge graph (KG)\footnote{Knowledge graph (KG) is a commonly used knowledge base structure that use graph topology to represent interlinked descriptions of entities. It has been widely used in many real world knowledge system applications like Google search engine \citep{Singhal2012KG}, Wikipedia encyclopedia, Facebook social networks, Apple's Siri, among many others. An example of knowledge graph used in Google search is illustrated in the left panel of Figure \ref{fig:kg_ex}. In Google's KG, it links many entities to famous people and store their relationships, like June 1723 as Adam Smith's birth month, and Kirkcaldy as his birth place. Such kind of \{subject, predicate, object\} triple structure like \{Adam Smith, birth place, Kirkcaldy\} is called \textit{RDF (Resource Description Framework) triples} in the terminology of knowledge graph. When a user search for Adam Smith's birth place, Google would provide webpages relevant to the entity ``Kirkcaldy'' based on this knowledge graph, even though those webpages may not be direct results of Google search algorithms like PageRank.} of the linkages between traditional economic variables and alternative data variables. We use advanced natural language processing (NLP) tools to extract such alternative data variables and their linkages from massive dataset the consists of academic literature and research reports. 
Specifically, we design an algorithm to extract from massive textual data (1) traditional variables of interest (like GDP, inflation rate, housing price, etc.), (2) alternative data variables (like electricity usage, migration flow, etc.), as well as (3) the relationships (positive correlation, negative correlation, etc.) among these variables. After some post-processing including resolving  coreferences, we build  a knowledge graph 
by starting with traditional variable of interests as the centers, and expanding in a step-wise fashion to include
the  relevant alternative variables. A very small subgraph of the resulted knowledge graph centered at \textit{inflation rate} is illustrated in the right panel of Figure \ref{fig:kg_ex}. This graph displays the conceptional relationship between \textit{inflation rate} and other  concepts discussed in
the massive textual data we study in this paper. Some of these linkages have already been well studied in the literature, for example  money supply will increase the inflation rate; the increase of benchmark interest rate may decrease the inflation rate. But there are also linkages that carry new information.  For example, we learn from some of these research reports that the increase of urban migration worker shortage may affect  inflation, through the increase of urban wages (upper right corner of Figure \ref{fig:kg_ex}).  \\

The knowledge graph we construct provides a new knowledge system for macroeconomics, and has many potential applications. 
One application we are particularly interested in is to formulate macroeconomics as a problem of reinforcement learning (RL) \citep{sutton2018reinforcement,silver2016mastering}.  A RL framework consists of the following essential components: the state space and the environment, the action space, the system dynamics and the reward functions. In this regard,  the knowledge graph of linkages among economic variables plays the role of the state space and the environment. In a companion paper \citep*{yang2020policy}, we construct a structured framework of economic policy targets and policy tools. The framework for the policy tools  plays the role of the action space in the RL framework, whereas the framework for the policy targets acts as the reward functions. In this paper, we also apply the knowledge graph of economic variables to a simple but more concrete task: variable selection in economic forecasting. Different from previous work using statistical tools to do variable selection, we use the knowledge graph as the prior knowledge to select variables for economic forecasting models. We will see that compared to statistical methods, the KG-based method achieves significantly higher forecasting accuracy, especially for long term forecasts.\\

This paper contributes to the following literature. First, we contribute to the literature on using big data and machine learning tools in macroeconomics \citep{mccracken2016fred, stock2016dynamic, giannone2008nowcasting,coulombe2019machine,fan2020factor}. Most of previous work apply or modify off-the-shelf machine learning tools to study high dimensional macroeconomic variables, while we stand out to design a new knowledge system for macroeconomics with big data. Second, we contribute to the literature on knowledge graph and knowledge extraction in science. Besides prominent applications in  industry, knowledge graph has also been used for knowledge extraction and knowledge representation in various scientific disciplines \citep{luan2018multi}, for example material science and physics. Last but not  the least, our application example contributes to the literature on variable selection and model reduction in economic forecasting with big data. Previous methods rely on statistical learning, including shrinkage methods like Lasso \citep{tibshirani1996regression}, Bayesian methods \citep*{sims1998bayesian,doan1984forecasting} and machine learning methods like factor models \citep*{fan2020recent} and autoencoder \citep*{goodfellow2016deep}. Our work provides a new method for variable selection, through a systematic treatment of existing human knowledge.\\ 


The remaining of the paper is organized as follows. In Section \ref{sec:data}, we introduce the textual data that we use to construct the knowledge graph, as well as the data we use in our application example. In Section \ref{sec:method}, we present the detailed algorithm for constructing the knowledge graph from massive textual data. We discuss the application of the knowledge graph on forecasting in Section \ref{sec:app}. In particular, we use it as the prior knowledge to do variable selection for economic forecasting. Finally we conclude with discussions on the future work.

\section{Data}\label{sec:data}
\subsection{Textual Data for Knowledge Graph Construction}
We build  the knowledge graph (KG) of linkages among economic variables by  extracting variables and their relationships from the
massive dataset that consists of previous research documents. In general, two different types of textual data are suited for our work: academic papers published by leading journals in economics, or research reports published by leading think tanks, consulting firms, asset management companies and other similar agencies (``industry research reports'' hereafter). In this paper, we use Chinese industry research reports as the major textual data source. We make this choice for the following reasons. First, most of those industry research reports focus on analyzing or forecasting the dynamics of aggregate variables, and it is always clearly stated what variables are studied in each report. Second, 
these reports mostly adopt the narrative approach \citep{shiller2017narrative} in research, which clearly state the logic chains of their analysis in narrative language, rather than in theoretical or quantitative models which is more common in English papers or industry reports. Thirdly, they are freely available and can be downloaded  from the WIND database\footnote{The WIND database is often referred to as the Chinese version of Bloomberg terminal, and is the major provider of macroeconomic and financial data and information in China. WIND also provides millions of industry research reports on the macroeconomy, industries or even individual public firms for download, which is not available on Bloomberg terminals.}. We download\footnote{
We downloaded all reports available on August 9, 2018 when we started this project.} all the industry research reports from the macroeconomic research section of the WIND database, and selected 846 of them that adopt the narrative approach to study certain variables of interest as the textual data for this paper.

\subsection{Traditional and Alternative Data for Economic Forecasting}
As an application of the knowledge graph we construct, we use it as the prior knowledge to select variables for economic forecasting models. In Section \ref{sec:app_ex}, we forecast China's monthly inflation and nominal investment time series using both traditional variable selection method and the KG-based method. For the traditional method, the input variables come from the standard Chinese monthly time series constructed by \cite*{higgins2015china} and \cite*{higgins2016forecasting}\footnote{The data is available at \url{https://www.frbatlanta.org/cqer/research/china-macroeconomy}, last accessed on August 21, 2020.}. For the period of 1996 to 2019, there are 12 monthly time series available: Real GDP,  Nominal Investment,	Nominal Consumption, M2, Nominal Imports, Nominal Exports, 7-Day Repo, Benchmark 1-year Deposit Rate, Nominal GDP, GDP Deflator, CPI and Investment Price.\\

For the KG-based method, we design model inputs under the guidance of the knowledge graph, and obtain as many data variables as we can from two major databases on Chinese economy: WIND and CEIC. The full list of alternative indicators we obtain are discussed in Section \ref{sec:app_ex}.

\section{Construction of Knowledge Graph}\label{sec:method}
\subsection{An Example: From Textual Data to Knowledge Graph}
Before presenting the detailed algorithm for extracting variables and their relations from the textual data, we illustrate the main idea with an example. The following paragraph is translated from a research report\footnote{The full text of this report (in Chinese) is available \url{https://www.nsd.pku.edu.cn/jzky/ceojjgcbgk/252274.htm. }, last accessed on August 15, 2020.} from a leading think tank in China.
It studies the dynamics of inflation rate using alternative data variables. We will explain how to build
 part of the knowledge graph from this paragraph step by step.\\

\noindent\fbox{%
    \parbox{\textwidth}{%
``Dr. Gao concluded that a long-term systematic migrant worker shortage began to appear in the Chinese migrant labor market around 2005, which greatly increased the growth rate of migrant workers' wages, resulted in the increase of food prices, and pushed up the increase in consumer price index, making the average level of  inflation probably 100 to 200 basis points higher.''
}}\\

In this example, we hope to extract all the economic variables\footnote{An economic variable is an object measurable with economic data. It could be traditional variables like GDP, inflation, and could also be alternative variables like migration flow, oil prices, among others.} and the relation keywords among those variables,  store them in \textit{RDF (Resource Description Framework)  triples} of \{variable 1, relation, variable 2\} format, and construct the knowledge graph from the \textit{RDF triples}. \\

As the first step, we  find the traditional variables of interest in this paragraph: consumer price index and inflation. These variables are the center nodes of the knowledge graphs. 
Next, we  find the alternative variables that are measurable with economic data: food prices, migrant workers' wages, migrant worker shortage. 
Thirdly, we find the key words linking these variables and the inflation variable: making...higher, push up, resulted in, increase. 
The extraction of all the \textbf{\underline{economic variables}} and the \textbf{\textit{relation keywords}} among those variables are highlighted in the box below.\\

\noindent\fbox{%
    \parbox{\textwidth}{%
``Dr. Gao concluded that a long-term systematic \textbf{\underline{migrant worker shortage}} began to appear in the Chinese migrant labor market around 2005, which greatly \textbf{\textit{increased}} the \textbf{\underline{growth rate of migrant workers' wages}}, \textbf{\textit{resulted in the increase}} of \textbf{\underline{food prices}}, and \textbf{\textit{pushed up}} the increase in \textbf{\underline{consumer price index}}, \textbf{\textit{making}} the average level of  \textbf{\underline{inflation}} probably 100 to 200 basis points \textbf{\textit{higher}}.''
}}\\

\begin{figure}[h!]
    \centering
    \includegraphics[width=0.75\textwidth]{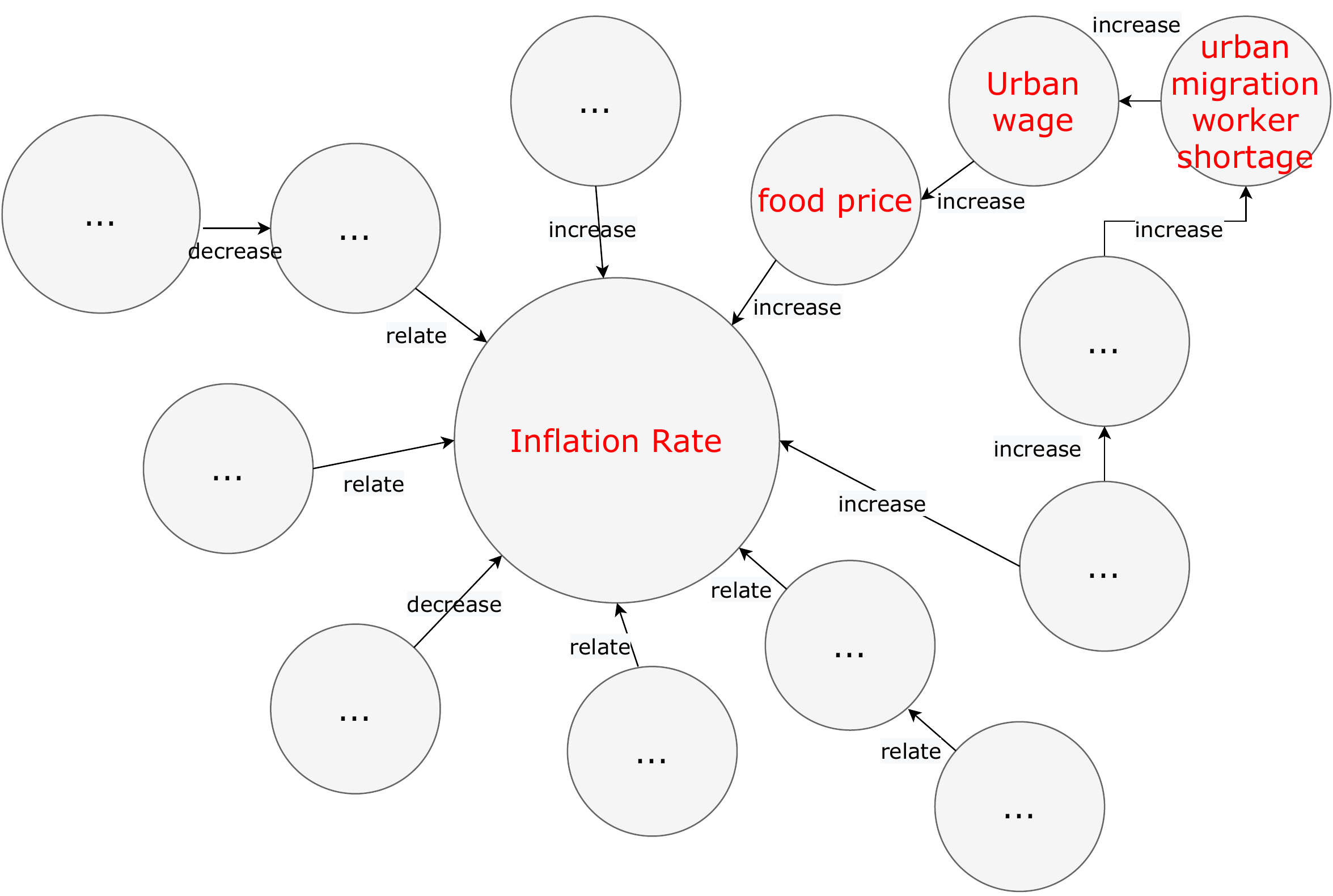}
    \caption{Knowledge Graph of Economic Variables: A Subgraph around the Inflation Rate}
    \label{fig:kg_ex1}
\end{figure}

We store all the results of the extraction process in \textit{RDF triples} of \{variable 1, relation, variable 2\} format: 
\begin{itemize}
    \item \{migrant worker shortage, increase, growth rate of migrant workers' wages\}
    \item \{growth rate of migrant workers' wages, resulted in the increase, food prices\}
    \item \{food prices, push up, consumer price index\}
    \item \{food prices, make higher, inflation\}
\end{itemize}
Later we will classify all the relation keywords into three classes: ``increase'', ``decrease'', ``neutral''. All the relation keywords here belong to the ``increase'' class. The last two \textit{RDF triples} are equivalent to each other, so the duplicates will be dropped when constructing the knowledge graph. 
This process gives rise  to the subgraph in the upper right corner of Figure \ref{fig:kg_ex1}.

\subsection{Construction Procedure}
Inspired by the example above, now we present the general procedures of constructing the knowledge graph of linkages among economic variables from the textual data.
\begin{enumerate}
    \item [Step 1.] Make a list of aggregate variables of interest, together with their variants. \\
    For this paper, we investigate the following variables (with their variants in the brackets): GDP (output, economic growth), Investment, Housing price (housing market, real estate market, real estate price), RMB Exchange Rate (RMB), Inflation (CPI).
    \item [Step 2.] Find all these aggregate variables and their variants in the documents with string matching.
    \item [Step 3.] For each aggregate variable detected in the documents, find all the other variables around it, as well as the relation among aggregate variables and other variables.
    \item [Step 4.] Represent all the variables and relations that have been  extracted with the typical \textit{RDF triple} structure in a knowledge graph.
    \item [Step 5.] Merge all the co-references and build the knowledge graph.
\end{enumerate}
Among all the procedures above, Steps 1, 2 and 4 are straightforward to implement, while Steps 3 and 5 are technically more challenging. We will discuss the challenges and how we address them in the next subsection.

\subsection{Main Challenges of Knowledge Graph Construction and Solutions}\label{challenge}
The most challenging steps of constructing the knowledge graph are the extraction of variable entities and relations, as well as co-reference resolution. \\

Economic variables, especially alternative variables, exhibit complicated semantic patterns that have not been investigated in classical entity recognition and co-reference resolution tasks \citep{ji2020survey, hogan2020knowledge}. Typically, entity recognition tasks include two steps: (1) identify regions of text that may correspond to entities, (2) categorize them into a predefined list of types (people, organization, location, among many others) \citep{ling2012fine}. Most of the literature take outputs from the first step as given, and focus on improving the classification work in the second step. However, in our problem, the first step is crucial and nontrivial, since economic variables are mostly multi-token entities with complicated semantic patterns. Examples of complicated variable entities include ``migration worker shortage'', ``growth rate of migration workers' wages'', ``processing firm registrations in China'', ``leverage rate of local government financing vehicles''. As a result, it is challenging to identify boundaries of the text that mentions a variable entity. For the second step, previous work mostly focuses on categorizing entities into names of people, organizations, locations, or other more detailed types labeles in large knowledge database like Freebase \citep{ling2012fine, hogan2020knowledge}, and are not suitable for classifying variable entities in our paper. Similarly, due to the complicated semantic patterns of economic variables, co-reference resolution, or entity disambiguation, is particularly challenging in our setting. \\

To address these challenges, we design a recursive weakly supervised learning algorithm \citep{zhou2018weaklysupervised} with some human involvement\footnote{To note, it is a common practice to involve direct contributions from human editors to construct knowledge graphs. Some prominent knowledge graphs are primarily constructed with human efforts \citep{hogan2020knowledge}.} to extract variable entities and entity relations from the textual data. In each iteration, we first identify regions of text that are most likely to have variable entities or relations, and then use human editor to select true variable entities or relations, and take the improved set into the next iteration. Our algorithm is similar to but different from the bootstrapping approach of entity recognition \citep{collins1999unsupervised,gupta2014improved}. To make this facilitation scheme work, we reduce the general notion of relation extraction to extracting only the ``relation keywords'' in the textual data. In terms of co-reference resolution, we also combine a similarity score measure and some human efforts to remove duplicates of entities.


\subsubsection{Entity Recognition: Variable Names and Relation Keywords}\label{app:extraction}
The weakly supervised learning algorithm to extract variable entities and relation keywords from the textual data works as follows:
\begin{enumerate}
    \item [Step 1.] Construct an initial \textit{set of economic variables} and an initial \textit{set of relation keywords}. The initial set of economic variables are from the macroeconomic database of WIND. The initial set of relation keywords are commonly seen relation words and phrases like increase, decrease, result in, push up, among others.
    \item [Step 2.] Using the current set of economic variables as training data, train a simple model to predict whether a phrase\footnote{A phrase is a N-gram object after word segmentation, where $N$ could be small integers with $N \leq 5$.} is a variable. With the model we can get a confidence metric for each phrase to be a variable. Human editors select true variables from those high-confident phrases to get an expanded \textit{variable set}.
    \item [Step 3.] Find sentences that contain many \textit{variables}, but very few \textit{relation keywords}. Then use the human editors to find the relation keywords in these sentences. Expand the \textit{set of relation keywords}.
    \item [Step 4.] Find sentences that contain relation keywords, but very few variables. Then use the human editors to find the variables in those sentences. Repeat Step 2 to expand the \textit{variable set}.
    \item [Step 5.] Repeat Steps 3 and 4, until we cannot find any new relation keywords or variables.
\end{enumerate}

\subsubsection{Co-reference Resolution}
We define the similarity score between two economic variable entities represented as two word vectors:
$$
\text{Sim}(u,v) = 1-\frac{u \cdot v}{\|u\|_{2}\|v\|_{2}}
$$
where $u, v$ are word vectors of the two variables. Based on this score, we get those economic variable entities with high similarity scores with each other. Human editors will determine the true duplicates from those high similarity pairs. Then we unify the names of those variable entities that are co-references with each other. The unification process is done by human editors to choose the best entity name that would show up in the knowledge graph. After the name unification, we remove the duplicated \textit{RDF triples}, and build the final knowledge graph based on the unique \textit{RDF triples} we get.

\subsection{Knowledge Graph Results}\label{sec:results}
\begin{figure}[h!]
    \centering
    \includegraphics[width=0.75\textwidth]{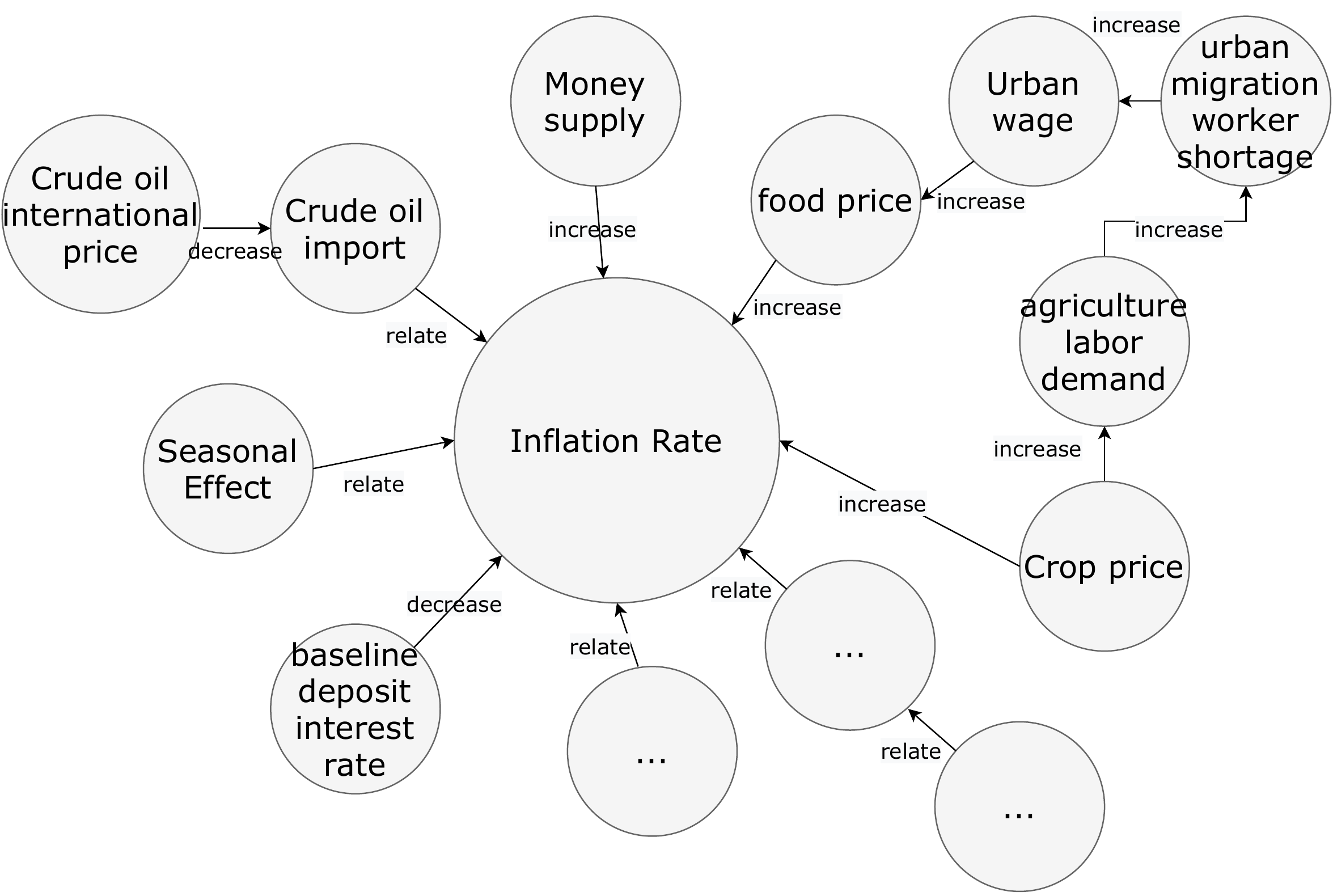}
    \caption{Knowledge Graph of Economic Variables: Inflation Rate}
    \label{fig:kg_results1}
\end{figure}

The knowledge graphs we construct are too big to be presented in the paper.
Here  we visualize some subgraphs to give an idea of what they look like. Figure \ref{fig:kg_results1} is the a small part of the knowledge graph centered at inflation rate, and subsumes the example we see in Figure \ref{fig:kg_ex1}.\\ 

Figure \ref{fig:kg_results1} exhibits several distinct features of the knowledge graph we construct. First, it links traditional variables of interest (like ``inflation rate'' here) to other variables, including traditional macroeconomic variables (like ``money supply'' at the top of Figure \ref{fig:kg_results1}) and alternative variables (like those variables in the upper right corner). Second, the variables are linked with relation keywords that can be mapped to three classes: increase (positive relation), decrease (negative relation), relate (neutral relation).
Third, the linkages might be one layer (like the linkage to money supply) or multiple layers (like the linkages in red). Multiple layers imply a logic chain to analyze the variables of interest. Finally, co-reference resolution is necessary to get the final knowledge graph, since there can be a large number of duplicated \textit{RDF triples} directly extracted from the textual data.\\

\begin{figure}[h!]
\begin{center}
  \subfloat[Knowledge Graph of Net Export]{\includegraphics[width=0.49\textwidth]{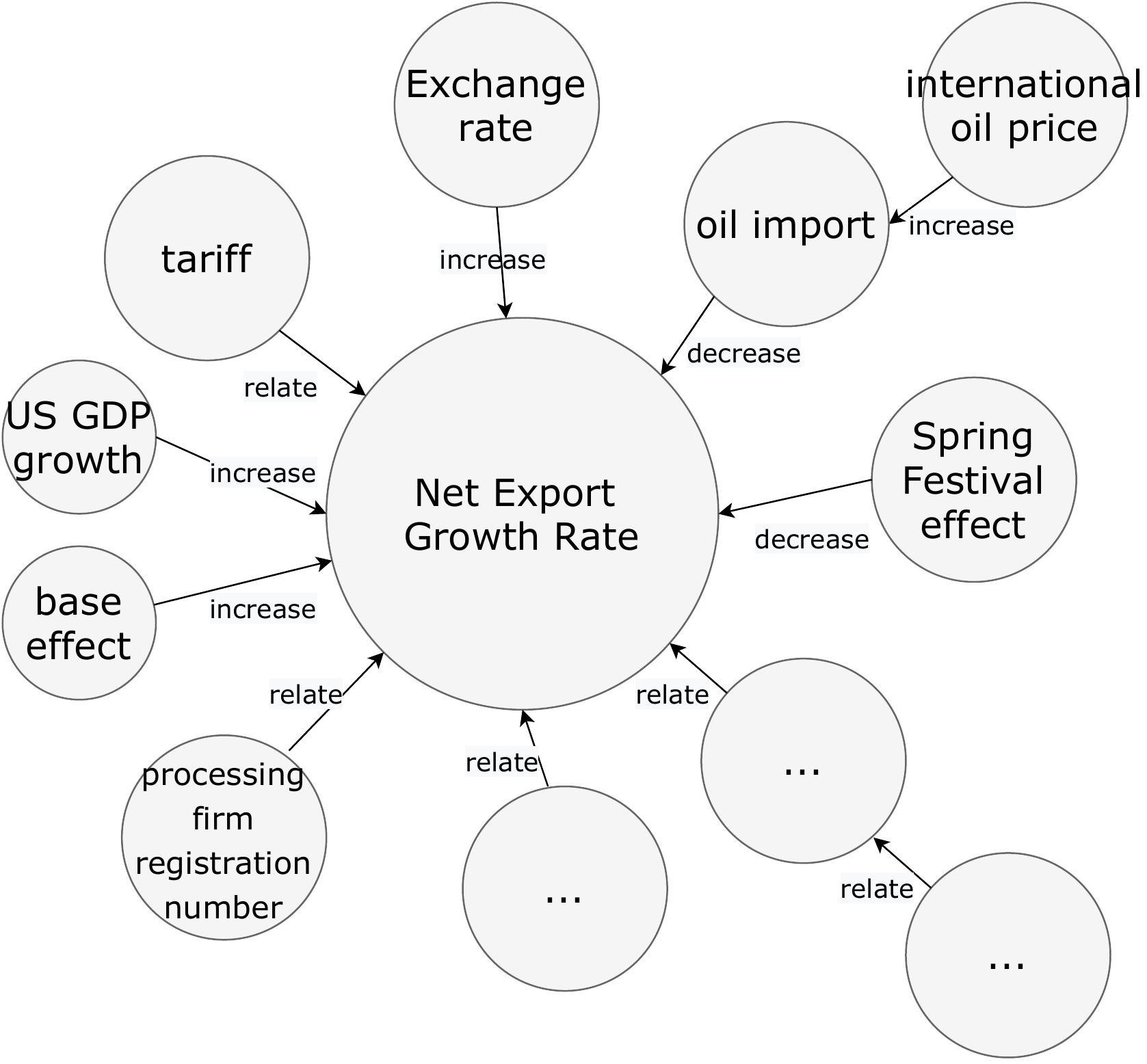}}
  \hfill
  \subfloat[Knowledge Graph of Housing Prices]{\includegraphics[width=0.49\textwidth]{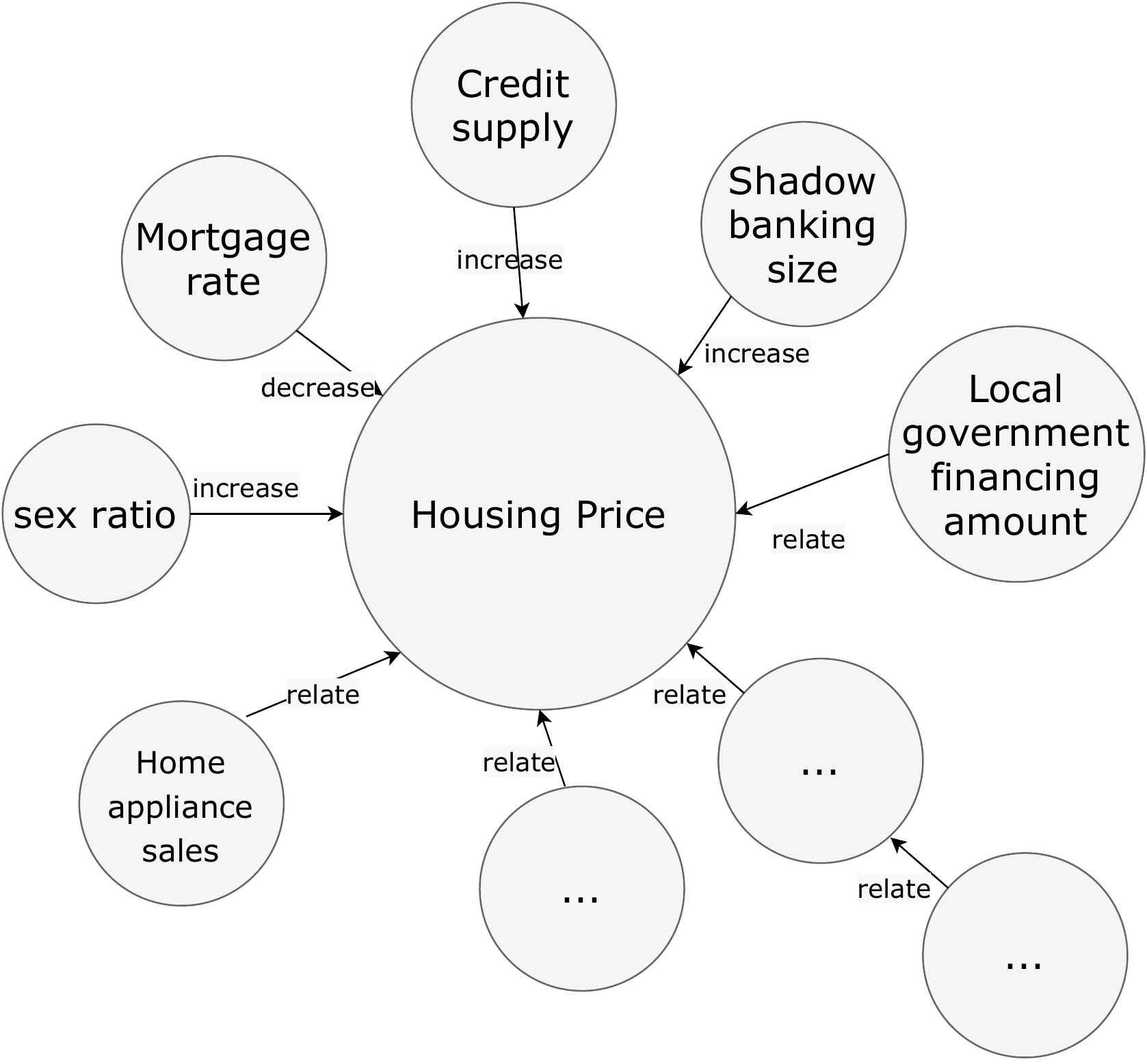}}
  \caption{Knowledge Graphs of Variable Linkages: More Examples}
    \label{fig:kg_results2}
  \end{center}
\end{figure}

More examples of the knowledge graphs are presented in Figure \ref{fig:kg_results2}. We would set up a website
to host the full results and we encourage people to join the effort to improve the knowledge graphs.

\section{Applications of Macroeconomic Knowledge Graph}\label{sec:app}
\subsection{Macroeconomics from a Reinforcement Learning Viewpoint}
The knowledge graph we construct provides a new knowledge system for macroeconomics. It  has many potential applications. One application we are particularly interested in is to study macroeconomics as a problem of reinforcement learning (RL) \citep{sutton2018reinforcement}, a framework
in machine learning for decision making problems.\\

An RL framework consists of the following essential components. First, the state space and the environment.
Here these should include the key variables that govern the evolution of the aggregate dynamics. The state space should consist of the variables that are affected by the actions of agents evolved, while the environment consists of  those that are not affected. Second, the action space, which consists of all the potential policy functions taken by agents (including households, firms and government agencies) in the economy. Third, the system dynamics, 
 the evolution law of the aggregate economy. 
 This can either be  model-based in which case we use a postulated dynamic model for the aggregate economy,
  or model-free in which case we rely purely on observational data. Fourth, the reward functions, here the utility functions of agents involved. In this regard,  the knowledge graph of linkages among economic variables helps to define the state space and the environment of the economy. In a companion paper \citep*{yang2020policy}, we construct a structured framework of economic policy tools and policy targets. The structure for the policy tools  plays the role of the action space in the RL framework, whereas the  policy targets act as the reward functions of government agencies. 
  Such a framework could potentially allow us to make maximum use of existing knowledge and data to study the macroeconomy.


 

\subsection{Example of Application: Knowledge Graph for Economic Forecasting}\label{sec:app_ex}

As a more concrete example, we apply the knowledge graph to the problem of variable selection in economic forecasting. Different from previous work using statistical tools to do variable selection \citep{tibshirani1996regression, zou2005regularization, sims1998bayesian, goodfellow2016deep, fan2020factor}, we combine statistical tools with the knowledge graph as the prior knowledge to select variables for economic forecasting models. 
Hopefully this will provide a new method for variable selection that uses systematically all existing human knowledge. \\

Here the task is to forecast China's monthly \textit{inflation rate} and \textit{nominal investment} time series from April 1996 to June 2019\footnote{The time range is chosen to have as many data observations as possible while still having a reasonably large number of input variables for both methods.}. 
For each $i = 1, 2, ..., 12$, we hope to build  a model to forecast the \textit{inflation rate} or \textit{nominal investment} $y_{t+i}$ in $i$ months ahead, with input variables from the past three months $\{\mathbf{X}_s\}_{s=t-3}^t$:
$$
y_{t+i} = f(\{\mathbf{X}_s\}_{s=t-3}^t)
$$

In  traditional statistical method, economists have little idea of what variables should be used as model inputs and what should not, so they typically find a standard dataset as input, and use statistical variable selection methods to estimate the model. In this spirit, we use the 12 time series variables\footnote{They are Real GDP,  Nominal Investment, Nominal Consumption, M2, Nominal Imports, Nominal Exports, 7-Day Repo, Benchmark 1-year Deposit Rate, Nominal GDP, GDP Deflator, CPI and Investment Price.} from the standard Chinese monthly time series constructed by \cite*{higgins2015china} and \cite*{higgins2016forecasting}, and use Lasso \citep{tibshirani1996regression} as the variable selection method.\\

For the KG-based method, we design model inputs with the guidance of the knowledge graph results in Section \ref{sec:results}, and obtain as many alternative data variables as we can from WIND and CEIC. Among all the alternative data variables\footnote{We convert all variables into the monthly frequency.} that are directly linked to \textit{inflation rate} in the knowledge graph, the following variables are available in the full model sample period (October 1996 to June 2019): CPI (historical data), GDP, benchmark 1-year deposit interest rate, benchmark 1-year loan interest rate, nationwide fiscal expenditure, urbanization rate, central government fiscal expenditure, share of manufacturing output in GDP, urban unemployment rate, worldwide GDP growth rate, M1 money supply, M2 money supply, USD/RMB exchange rate, crude oil production, raw coal production, copper production, raw coal production, Non-ferrous metal production, OPEC Basket Price, crude oil import amount, raw coal import amount, copper import amount, steel import amount, Spring Festival dummy, National Day Festival dummy. Among all the alternative data variables that are directly linked to \textit{nominal investment} in the knowledge graph, the following variables are available in the full model sample period (April 1996 to June 2019): nominal investment (historical data), GDP, benchmark 1-year deposit interest rate, benchmark 1-year loan interest rate, nationwide fiscal expenditure, central government fiscal expenditure, refinery capacity, metal smelter capacity, economic policy uncertainty (EPU) index, tariff income, tax income, stock market return, stock market volatility, dummy variable for the reform of replacing business tax with value-added tax, dummy variable for central leadership transition, M1 money supply, M2 money supply, Spring Festival dummy, National Day Festival dummy. These variables are used in Lasso regression to give the KG-based model prediction.\\

\begin{figure}[h!]
\begin{center}
  \subfloat[MAPE of Inflation Forecasting]{\includegraphics[width=0.499\textwidth]{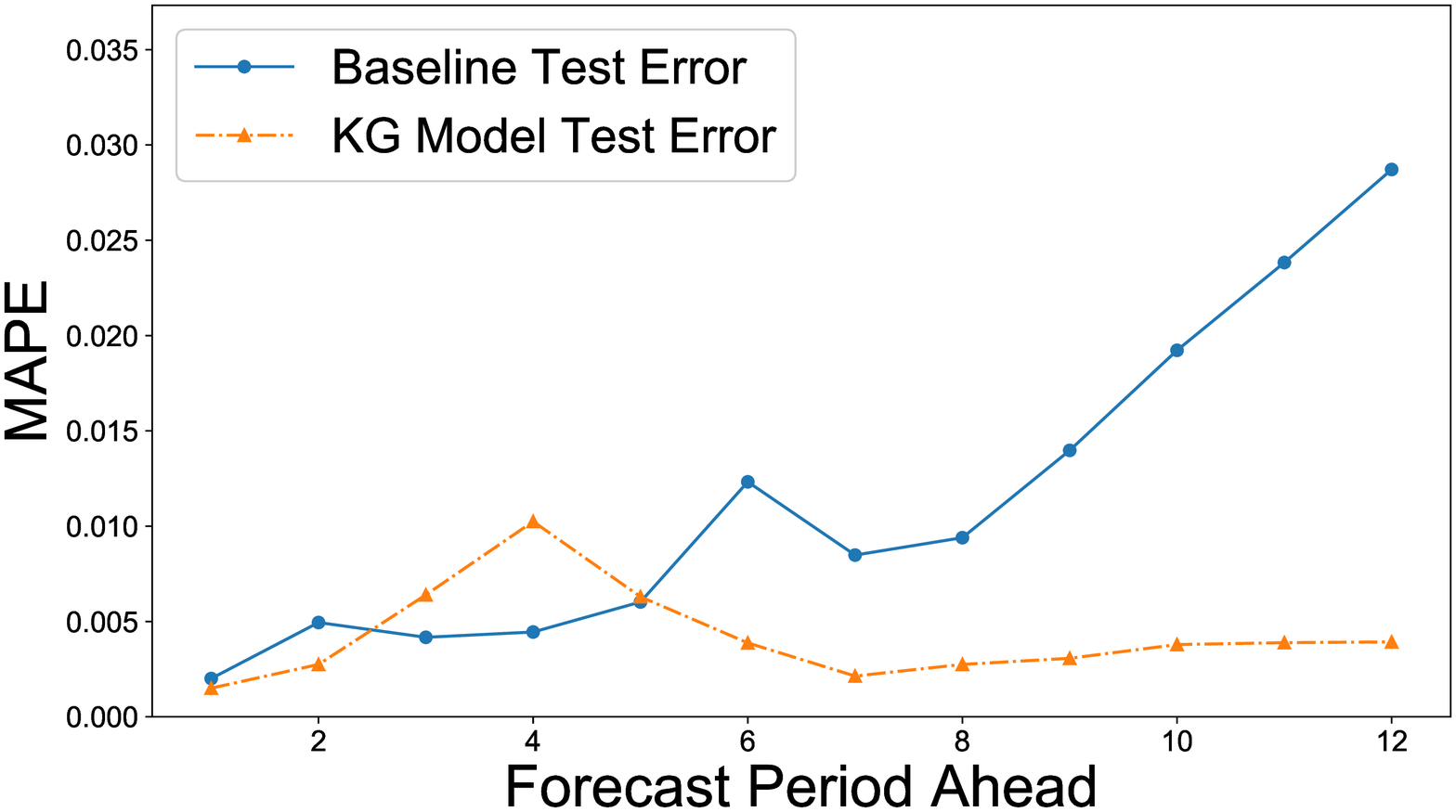}}
  \hfill
  \subfloat[RMSE of Inflation Forecasting]{\includegraphics[width=0.499\textwidth]{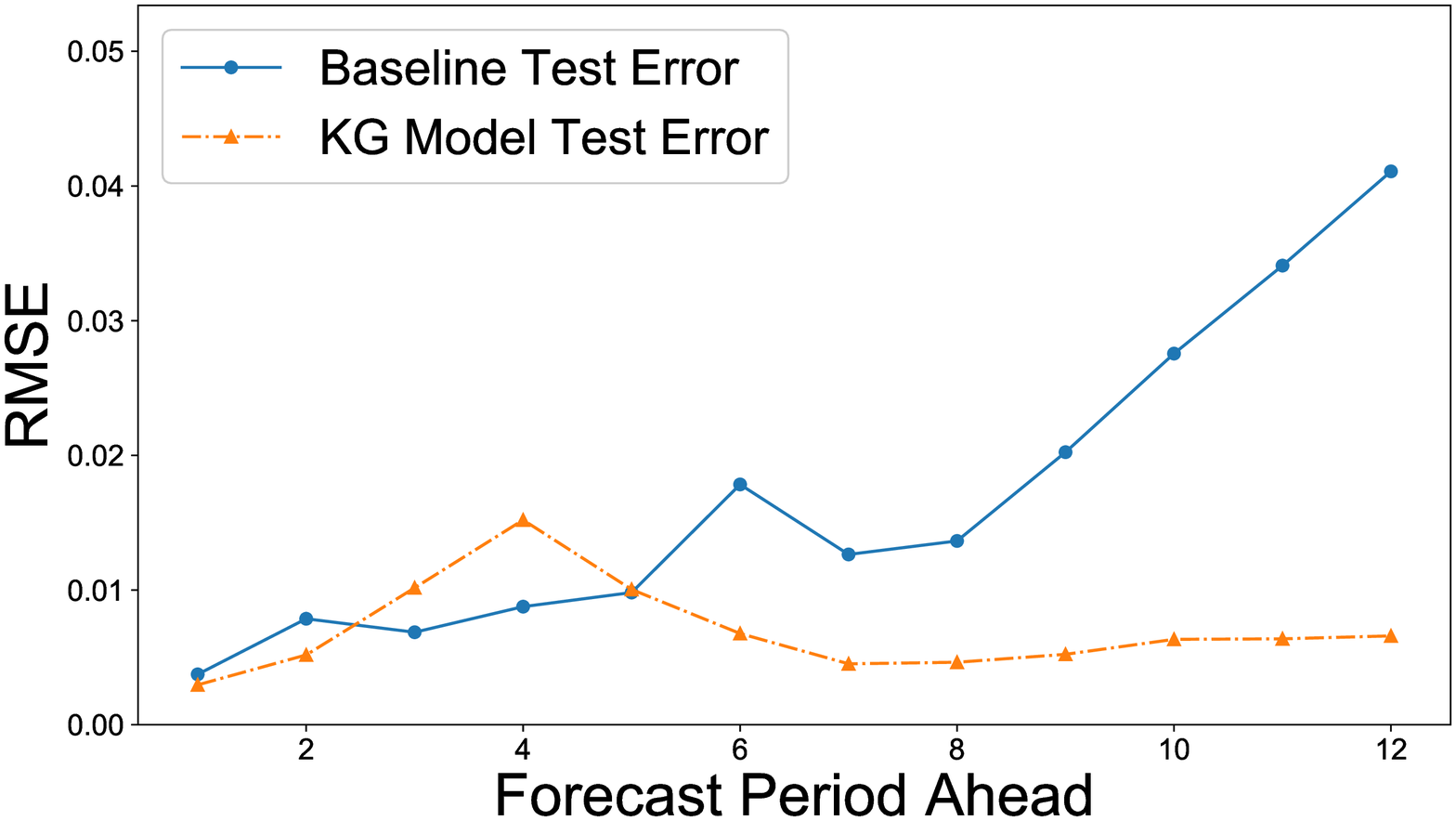}}\\
  \hfill
  \subfloat[MAPE of Investment Forecasting]{\includegraphics[width=0.499\textwidth]{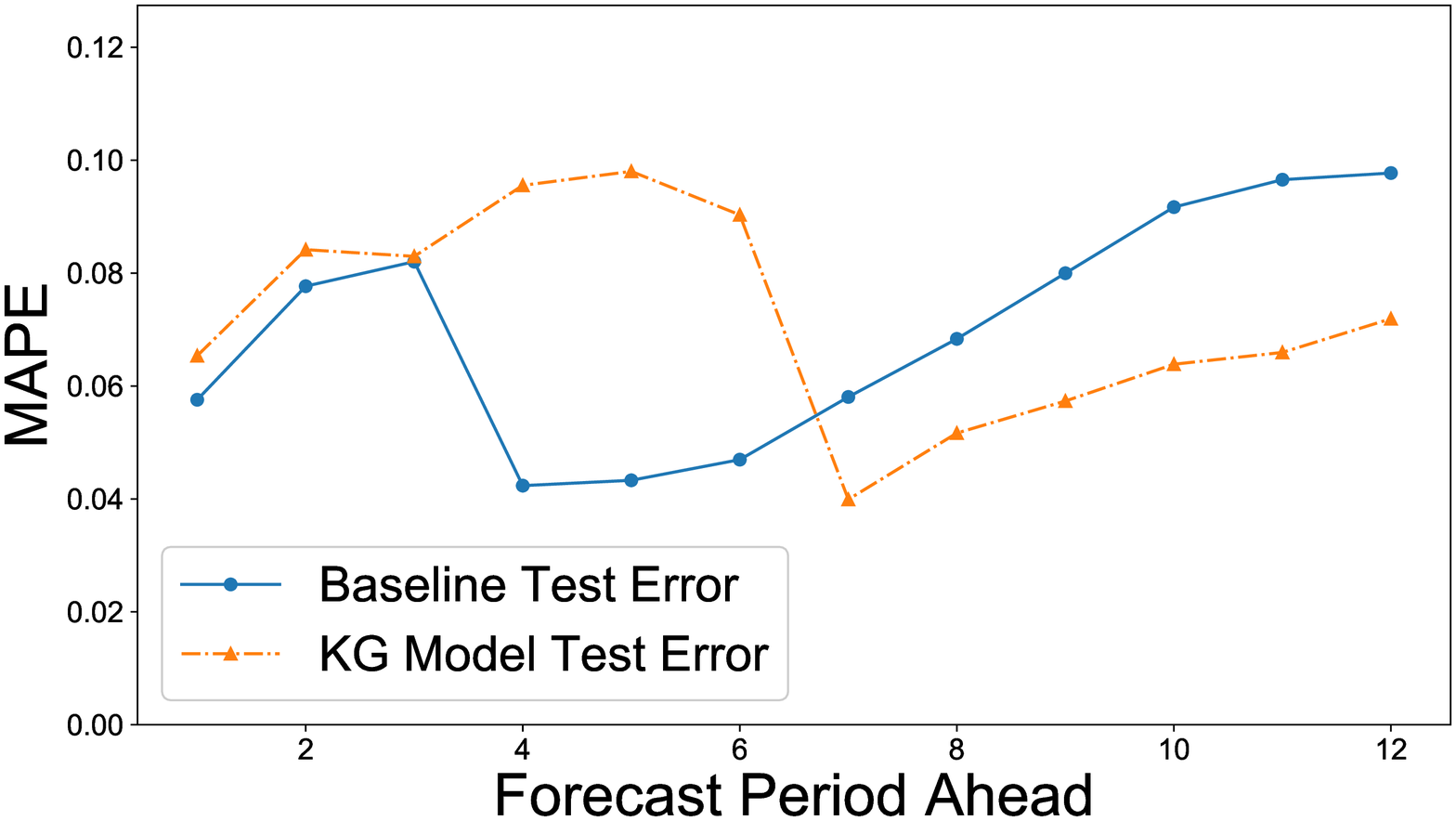}}
  \hfill
  \subfloat[RMSE of Investment Forecasting]{\includegraphics[width=0.499\textwidth]{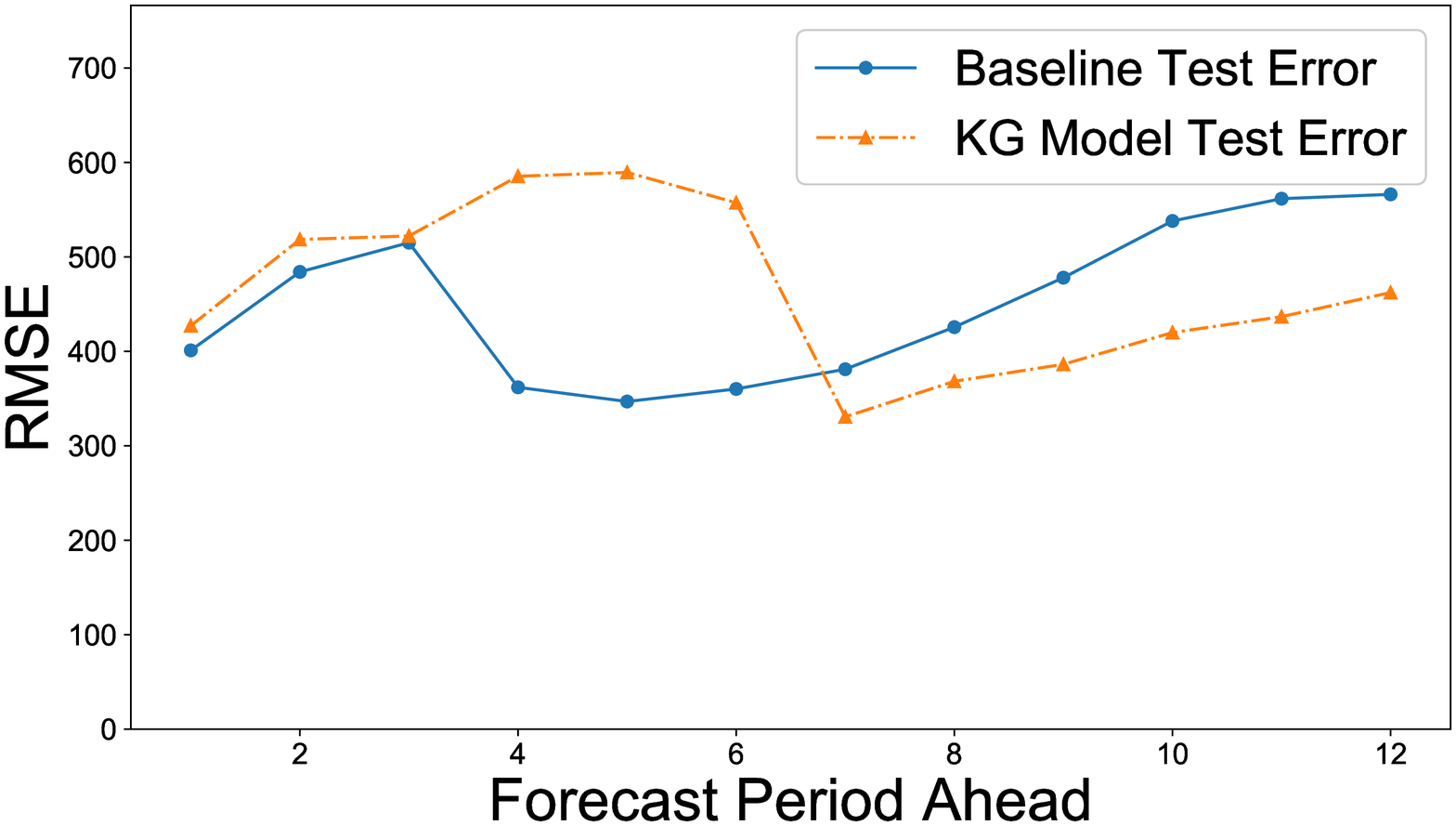}} \caption{Forecasting Errors of Baseline Model vs. KG-Based Model}
    \label{fig:compare}
  \end{center}
\end{figure}

For different forecasting periods  (from one month to 12 months), the forecasting errors on the test sets 
for both the baseline model and KG-based model are presented in Figure \ref{fig:compare}. We report the mean absolute percentage error (MAPE) in the left panels, and the root mean squared error (RMSE) in the right panels, and the results are qualitatively the same. For inflation forecasting (upper panels), compared to the baseline model, the KG-based model achieves higher forecasting accuracy in general. In short term forecasting (within five months), the forecasting errors for both models  are comparable to each other, and the baseline model even outperforms the KG-based model in some horizons.  However, in long run forecasting, the performance of the baseline model gets worse, while the KG-based model achieves a stable and much higher accuracy than the baseline method. Similar arguments also hold for nominal investment forecasting (bottom panels in Figure \ref{fig:compare}). \\

The general trend revealed by Figure \ref{fig:compare} are consistent with our expectation that short term forecasting relies more on data, while long term forecasting relies more on capturing the underlying logic in the problem.
The baseline model is more of a pure data-driven model, whereas the KG-based model tries to capture the underlying logic. The better long term performance of the KG-based model serves as a confirmation that the relationships described in the knowledge graph correctly represents the true logic of the economic system under investigation.\\
\begin{table}[h!]
\resizebox{1.1\textwidth}{!}{%
\begin{tabular}{|c|c|c|c|c|c|c|c|c|c|c|c|c|}
\hline
Periods ahead & 1 & 2 & 3 & 4 & 5 & 6 & 7 & 8 & 9 & 10 & 11 & 12 \\ \hline
P-value: inflation & 0.1102 & 0.0157 & 0.0248 & 0.0971 & 0.3213 & 0.0000 & 0.0002 & 0.0099 & 0.0020 & 0.0002 & 0.0000 & 0.0002 \\ \hline
P-value: investment & 0.1128 & 0.0799 & 0.5677 & 0.0941 & 0.0162 & 0.0406 & 0.0005 & 0.2295 & 0.0426 & 0.0064 & 0.0104 & 0.0693 \\ \hline
\end{tabular}}
\caption{The p-values of Diebold-Mariano Tests}
\label{tab:comp}
\end{table}

To check the significance of these comparisons, we perform the Diebold-Mariano (DM) test \citep*{diebold1995comparing,harvey1997testing} for both forecasting problems. We report the p-values of the DM tests for different forecasting periods ahead in Table \ref{tab:comp}. Most of the comparison results are significant, which again confirms our findings.

\section{Conclusion}
In the age of big data,  traditional knowledge system of macroeconomics that built on interactions among a small number of variables is faced with severe challenges. In this paper, we develop an approach to build a knowledge graph (KG) of the linkages between traditional economic variables and massive alternative big data variables. We extract these variables and linkages by applying advanced natural language processing (NLP) tools on the massive dataset that consists of academic literature and research reports. \\

The knowledge graph we construct has many potential applications. In this paper, we use it as the prior knowledge to select variables for economic forecasting models in macroeconomics. Compared to statistical variable selection methods, KG-based method achieves lower forecasting errors, especially for long run forecasts. In this particular example, we only make use of the list of variables around the variable of interest (like inflation) in the knowledge graph, rather than the multi-layer graphical structure of knowledge embedding. Future research can further investigate how to incorporate this structure into the statistical model, possibly in the form of model structure restrictions. Future work may also investigate other potential applications of this new knowledge system, like the reinforcement learning framework of macroeconomics we discuss briefly in this paper.

\singlespacing
\bibliography{ChinaDevo}

\begin{thebibliography}{30}
\newcommand{\enquote}[1]{``#1''}
\providecommand{\natexlab}[1]{#1}
\providecommand{\url}[1]{\texttt{#1}}
\providecommand{\urlprefix}{URL }
\providecommand{\bibAnnoteFile}[1]{%
  \IfFileExists{#1}{\begin{quotation}\noindent\textsc{Key:} #1\\
  \textsc{Annotation:}\ \input{#1}\end{quotation}}{}}
\providecommand{\bibAnnote}[2]{%
  \begin{quotation}\noindent\textsc{Key:} #1\\
  \textsc{Annotation:}\ #2\end{quotation}}

\bibitem[{Christiano et~al.(2005)Christiano, Eichenbaum, and
  Evans}]{christiano2005nominal}
Christiano, Lawrence~J, Martin Eichenbaum, and Charles~L Evans (2005),
  \enquote{Nominal rigidities and the dynamic effects of a shock to monetary
  policy.} \emph{Journal of Political Economy}, 113, 1--45.
\bibAnnoteFile{christiano2005nominal}

\bibitem[{Collins and Singer(1999)}]{collins1999unsupervised}
Collins, Michael and Yoram Singer (1999), \enquote{Unsupervised models for
  named entity classification.} In \emph{1999 Joint SIGDAT Conference on
  Empirical Methods in Natural Language Processing and Very Large Corpora}.
\bibAnnoteFile{collins1999unsupervised}

\bibitem[{Coulombe et~al.(2019)Coulombe, Leroux, Stevanovic, and
  Surprenant}]{coulombe2019machine}
Coulombe, Philippe~Goulet, Maxime Leroux, Dalibor Stevanovic, and St{\'e}phane
  Surprenant (2019), \enquote{How is machine learning useful for macroeconomic
  forecasting?} Technical report, CIRANO.
\bibAnnoteFile{coulombe2019machine}

\bibitem[{Diebold and Mariano(1995)}]{diebold1995comparing}
Diebold, Francis~X and Robert~S Mariano (1995), \enquote{Comparing predictive
  accuracy.} \emph{Journal of Business \& economic statistics}, 20, 134--144.
\bibAnnoteFile{diebold1995comparing}

\bibitem[{Doan et~al.(1984)Doan, Litterman, and Sims}]{doan1984forecasting}
Doan, Thomas, Robert Litterman, and Christopher Sims (1984),
  \enquote{Forecasting and conditional projection using realistic prior
  distributions.} \emph{Econometric reviews}, 3, 1--100.
\bibAnnoteFile{doan1984forecasting}

\bibitem[{Fan et~al.(2020{\natexlab{a}})Fan, Ke, and Wang}]{fan2020factor}
Fan, Jianqing, Yuan Ke, and Kaizheng Wang (2020{\natexlab{a}}),
  \enquote{Factor-adjusted regularized model selection.} \emph{Journal of
  Econometrics}.
\bibAnnoteFile{fan2020factor}

\bibitem[{Fan et~al.(2020{\natexlab{b}})Fan, Li, and Liao}]{fan2020recent}
Fan, Jianqing, Kunpeng Li, and Yuan Liao (2020{\natexlab{b}}), \enquote{Recent
  developments on factor models and its applications in econometric learning.}
  \emph{arXiv preprint arXiv:2009.10103}.
\bibAnnoteFile{fan2020recent}

\bibitem[{Giannone et~al.(2008)Giannone, Reichlin, and
  Small}]{giannone2008nowcasting}
Giannone, Domenico, Lucrezia Reichlin, and David Small (2008),
  \enquote{Nowcasting: The real-time informational content of macroeconomic
  data.} \emph{Journal of Monetary Economics}, 55, 665--676.
\bibAnnoteFile{giannone2008nowcasting}

\bibitem[{Goodfellow et~al.(2016)Goodfellow, Bengio, and
  Courville}]{goodfellow2016deep}
Goodfellow, Ian, Yoshua Bengio, and Aaron Courville (2016), \emph{Deep
  learning}. MIT press.
\bibAnnoteFile{goodfellow2016deep}

\bibitem[{Gupta and Manning(2014)}]{gupta2014improved}
Gupta, Sonal and Christopher~D Manning (2014), \enquote{Improved pattern
  learning for bootstrapped entity extraction.} In \emph{Proceedings of the
  Eighteenth Conference on Computational Natural Language Learning}, 98--108.
\bibAnnoteFile{gupta2014improved}

\bibitem[{Harvey et~al.(1997)Harvey, Leybourne, and
  Newbold}]{harvey1997testing}
Harvey, David, Stephen Leybourne, and Paul Newbold (1997), \enquote{Testing the
  equality of prediction mean squared errors.} \emph{International Journal of
  forecasting}, 13, 281--291.
\bibAnnoteFile{harvey1997testing}

\bibitem[{Higgins et~al.(2016)Higgins, Zha, and Zhong}]{higgins2016forecasting}
Higgins, Patrick, Tao Zha, and Wenna Zhong (2016), \enquote{Forecasting
  {C}hina's economic growth and inflation.} \emph{China Economic Review}, 41,
  46--61.
\bibAnnoteFile{higgins2016forecasting}

\bibitem[{Higgins and Zha(2015)}]{higgins2015china}
Higgins, Patrick~C and Tao Zha (2015), \enquote{{C}hina’s macroeconomic time
  series: Methods and implications.} \emph{Unpublished Manuscript, Federal
  Reserve Bank of Atlanta}.
\bibAnnoteFile{higgins2015china}

\bibitem[{Hogan et~al.(2020)Hogan, Blomqvist, Cochez, d'Amato, de~Melo,
  Gutierrez, Gayo, Kirrane, Neumaier, Polleres et~al.}]{hogan2020knowledge}
Hogan, Aidan, Eva Blomqvist, Michael Cochez, Claudia d'Amato, Gerard de~Melo,
  Claudio Gutierrez, Jos{\'e} Emilio~Labra Gayo, Sabrina Kirrane, Sebastian
  Neumaier, Axel Polleres, et~al. (2020), \enquote{Knowledge graphs.}
  \emph{arXiv preprint arXiv:2003.02320}.
\bibAnnoteFile{hogan2020knowledge}

\bibitem[{Ji et~al.(2020)Ji, Pan, Cambria, Marttinen, and Yu}]{ji2020survey}
Ji, Shaoxiong, Shirui Pan, Erik Cambria, Pekka Marttinen, and Philip~S Yu
  (2020), \enquote{A survey on knowledge graphs: Representation, acquisition
  and applications.} \emph{arXiv preprint arXiv:2002.00388}.
\bibAnnoteFile{ji2020survey}

\bibitem[{Ling and Weld(2012)}]{ling2012fine}
Ling, Xiao and Daniel~S Weld (2012), \enquote{Fine-grained entity recognition.}
  In \emph{Proceedings of the Twenty-Sixth AAAI Conference on Artificial
  Intelligence}, 94--100.
\bibAnnoteFile{ling2012fine}

\bibitem[{Luan et~al.(2018)Luan, He, Ostendorf, and Hajishirzi}]{luan2018multi}
Luan, Yi, Luheng He, Mari Ostendorf, and Hannaneh Hajishirzi (2018),
  \enquote{Multi-task identification of entities, relations, and coreference
  for scientific knowledge graph construction.} In \emph{Proceedings of the
  2018 Conference on Empirical Methods in Natural Language Processing (EMNLP)},
  3219--3232.
\bibAnnoteFile{luan2018multi}

\bibitem[{McCracken and Ng(2016)}]{mccracken2016fred}
McCracken, Michael~W and Serena Ng (2016), \enquote{{FRED-MD}: A monthly
  database for macroeconomic research.} \emph{Journal of Business \& Economic
  Statistics}, 34, 574--589.
\bibAnnoteFile{mccracken2016fred}

\bibitem[{Shiller(2017)}]{shiller2017narrative}
Shiller, Robert~J (2017), \enquote{Narrative economics.} \emph{American
  Economic Review}, 107, 967--1004.
\bibAnnoteFile{shiller2017narrative}

\bibitem[{Silver et~al.(2016)Silver, Huang, Maddison, Guez, Sifre, Van
  Den~Driessche, Schrittwieser, Antonoglou, Panneershelvam, Lanctot
  et~al.}]{silver2016mastering}
Silver, David, Aja Huang, Chris~J Maddison, Arthur Guez, Laurent Sifre, George
  Van Den~Driessche, Julian Schrittwieser, Ioannis Antonoglou, Veda
  Panneershelvam, Marc Lanctot, et~al. (2016), \enquote{Mastering the game of
  go with deep neural networks and tree search.} \emph{Nature}, 529, 484--489.
\bibAnnoteFile{silver2016mastering}

\bibitem[{Sims(1972)}]{sims1972money}
Sims, Christopher~A (1972), \enquote{Money, income, and causality.} \emph{The
  American Economic Review}, 62, 540--552.
\bibAnnoteFile{sims1972money}

\bibitem[{Sims and Zha(1998)}]{sims1998bayesian}
Sims, Christopher~A and Tao Zha (1998), \enquote{Bayesian methods for dynamic
  multivariate models.} \emph{International Economic Review}, 949--968.
\bibAnnoteFile{sims1998bayesian}

\bibitem[{Singhal(2012)}]{Singhal2012KG}
Singhal, Amit (2012), \enquote{Introducing the knowledge graph: things, not
  strings.} \emph{Google Blog},
  \urlprefix\url{https://www.blog.google/products/search/introducing-knowledge-graph-things-not/}.
\bibAnnoteFile{Singhal2012KG}

\bibitem[{Smets and Wouters(2007)}]{smets2007shocks}
Smets, Frank and Rafael Wouters (2007), \enquote{Shocks and frictions in us
  business cycles: A bayesian dsge approach.} \emph{American Economic Review},
  97, 586--606.
\bibAnnoteFile{smets2007shocks}

\bibitem[{Stock and Watson(2016)}]{stock2016dynamic}
Stock, James~H and Mark~W Watson (2016), \enquote{Dynamic factor models,
  factor-augmented vector autoregressions, and structural vector
  autoregressions in macroeconomics.} In \emph{Handbook of macroeconomics},
  volume~2, 415--525, Elsevier.
\bibAnnoteFile{stock2016dynamic}

\bibitem[{Sutton and Barto(2018)}]{sutton2018reinforcement}
Sutton, Richard~S and Andrew~G Barto (2018), \emph{Reinforcement learning: An
  introduction}. MIT press.
\bibAnnoteFile{sutton2018reinforcement}

\bibitem[{Tibshirani(1996)}]{tibshirani1996regression}
Tibshirani, Robert (1996), \enquote{Regression shrinkage and selection via the
  lasso.} \emph{Journal of the Royal Statistical Society: Series B
  (Methodological)}, 58, 267--288.
\bibAnnoteFile{tibshirani1996regression}

\bibitem[{Yang et~al.(2020)Yang, Shao, and E}]{yang2020policy}
Yang, Yucheng, Zihao Shao, and Weinan E (2020), \enquote{Understanding
  {C}hina's policy making via machine learning.} Technical report, Princeton
  University.
\bibAnnoteFile{yang2020policy}

\bibitem[{Zhou(2018)}]{zhou2018weaklysupervised}
Zhou, Zhi-Hua (2018), \enquote{A brief introduction to weakly supervised
  learning.} \emph{National Science Review}, 5, 44--53.
\bibAnnoteFile{zhou2018weaklysupervised}

\bibitem[{Zou and Hastie(2005)}]{zou2005regularization}
Zou, Hui and Trevor Hastie (2005), \enquote{Regularization and variable
  selection via the elastic net.} \emph{Journal of the royal statistical
  society: series B (statistical methodology)}, 67, 301--320.
\bibAnnoteFile{zou2005regularization}

\end{thebibliography}

\end{document}